\documentclass[aps,prb,twocolumn,showpacs,tightenlines,amsmath]{revtex4}
\usepackage{graphics}
\usepackage{epsfig}
\begin{document}
\title{\Large \bf{
Adiabatic quantum pump in the presence of external ac voltages}} 
\author{
M. Moskalets$^{1}$ 
and
M. B\"uttiker$^{2}$
}
\affiliation{
         $^1$Department of Metal and Semiconductor Physics,\\
        National Technical University "Kharkov Polytechnical Institute",
        61002 Kharkov, Ukraine\\
        $^2$D\'epartement de Physique Th\'eorique, Universit\'e de Gen\`eve,
        CH-1211 Gen\`eve 4, Switzerland\\}

\date\today
 
\begin{abstract}
We investigate a quantum pump 
which in addition to its dynamic pump parameters is subject
to oscillating 
external potentials applied to the contacts of the sample. 
Of interest is the rectification of the ac currents flowing through 
the mesoscopic scatterer and their interplay with the quantum pump 
effect. 
We calculate the adiabatic dc current arising under the simultaneous 
action of both the quantum pump effect and classical rectification.
In addition to two known terms we find a third novel contribution 
which arises from the interference of the ac currents generated by the 
external potentials and the ac currents generated by the pump. 
The interference contribution renormalizes both the quantum pump effect 
and the ac rectification effect. Analysis of this interference effect 
requires a calculation of the Floquet scattering matrix beyond the 
adiabatic approximation based on the frozen scattering matrix alone. 
The results permit us to find the instantaneous current. In addition 
to the current generated by the oscillating potentials, and the ac current 
due to the variation of the charge of the frozen scatterer, there is 
a third contribution which represents the ac currents generated by an 
oscillating scatterer. 
We argue that the resulting pump effect can be viewed as a quantum 
rectification of the instantaneous ac currents generated by the 
oscillating scatterer. 
These instantaneous currents are an intrinsic property of a 
nonstationary scattering process.
\end{abstract}


\maketitle

\small 

\section{Introduction}
\label{intro}

Dynamical transport in mesoscopic structures attracts 
presently considerable attention 
\cite{BTP94}$^{-}$\cite{DMH03}.
In particular the possibility to vary 
several parameters at the same frequency but different phases 
\cite{SMCG99}
of a mesoscopic system opens up 
new prospects for the investigation of quantum transport.
Applying two slowly oscillating potentials 
at frequency $\omega$ with fixed 
phase lag $\Delta\varphi$  
to a mesoscopic conductor connected to reservoirs having equal 
electrochemical potentials one can generate an adiabatic dc current  
\begin{equation}
\label{Eq1}
I_{dc} \sim \omega \sin(\Delta\varphi).
\end{equation}
\noindent
Such a current was measured experimentally \cite{SMCG99}. However
the precise origin of the measured current is still unclear. 
At least two mechanisms considered in the literature can contribute to 
the experimentally measured current.
First, there exists a {\it quantum pump effect} 
\cite{SMCG99,Brouwer98}$^{-}$\cite{EWAK03}
which is due to quantum-mechanical interference and dynamical 
breaking of time-reversal 
invariance. 
Second, there also exists a {\it rectification of ac 
currents} by the oscillating scatterer
\cite{Brouwer01,PB01,DMH03}
if it is part of an external circuit with non-zero impedance.
Closely related to this 
second effect is a pump in the presence of inelastic
scattering: in addition to the externally driven pump parameters, 
inelastic scattering leads to an effective oscillating (electro-)chemical 
potential of the pump which acts like an additional pump parameter \cite{MB01}.

\begin{figure}[b]
  \vspace{3mm}
  \centerline{
   \epsfxsize 7cm
   \epsffile{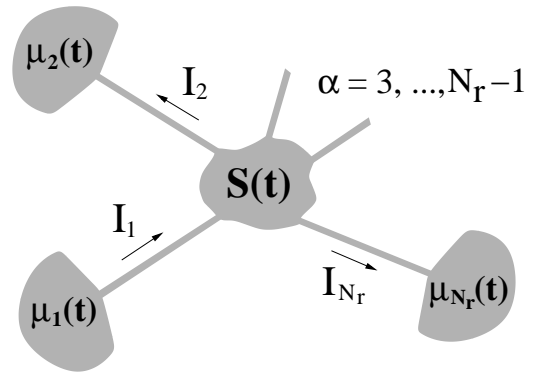}
             }
  \vspace{3mm}
  \nopagebreak
    \caption{
A mesoscopic pump with scattering matrix $S(t)$ oscillating 
with frequency $\omega$
is coupled to $N_r$ reservoirs with electrochemical potentials 
$\mu_{\alpha}(t)$ 
oscillating with the same frequency $\omega$. 
A quantum pump effect and a classical rectification 
effect together result in dc currents $I_{\alpha}$ 
flowing through the scatterer. The full current is time-dependent 
and is needed to characterize pumps in a non-zero impedance
external circuit. 
}
\label{fig1}
\end{figure}

The aim of the present paper is to investigate both these mechanisms on 
the same footing. To this end we consider a phase coherent oscillating 
scatterer coupled to reservoirs with oscillating potentials (see Fig. 1). 
We will show that in general the above mentioned mechanisms 
do not simply add but interfere between 
themselves. This leads to a renormalization of 
both the quantum pump effect as well as the rectification effect 
in the total dc current.

Theoretically quantum pumps have been investigated mostly under 
the (implicit) condition that the external circuit exhibits zero-impedance.
The work of Brouwer \cite{Brouwer01}, Polianski and Brouwer \cite{PB01},
the work on inelastic 
scattering \cite{MB01} mentioned already 
and the recent work of Martinez-Mares, Lewenkopf
and Mucciolo \cite{MMLM} represent the few exceptions. 
In reality the 
zero-impedance condition seems never exactly fulfilled. Coupling  
an oscillating gate voltage to a scatterer leads,
due to the long range nature of the Coulomb interaction, 
effectively to oscillating voltages at all terminals \cite{DMH03}.
In addition, in experiments, the pump is investigated with an impedance 
in series  with the oscillating scatterer. Furthermore, a voltage probe, 
to maintain zero current in the presence of pumping, in effect generates
an oscillating potential which acts back on the pump \cite{MB01}. 
Therefore, for theory to make 
contact with experiment, it is necessary to consider the effect of oscillating 
voltages at the contacts of the conductor.

The paper is organized as follows.
In Sec.\ref{GA} we develop the Floquet scattering matrix 
approach for ac quantum transport
through the nonstationary (oscillating) scatterer in the presence 
of oscillating reservoir potentials. A full theory requires even 
to first order in frequency an investigation of 
non-adiabatic corrections to the adiabatic (frozen) scattering matrix.
These corrections are discussed in Sec.\ref{AA}.
The current to linear order in the reservoir potentials is calculated 
in Sec.\ref{LR}. We illustrate the results for a simple 
one-channel scatterer with two contacts. In Sec.\ref{FV} we
present a general expression for the current valid for finite 
potentials. Sec. \ref{IC} gives the expression for 
the instantaneous current.

\section{General approach}
\label{GA}

For simplicity we consider here the mesoscopic sample, the pump, connected 
to $N_{r}$ reservoirs via single channel leads Fig.{\ref{fig1}}.
We are interested in the dc and ac currents flowing in the system 
if this system is subject to 
a cyclic evolution with period ${\cal{T}}$. 
The general situation we want to consider 
admits the scatterer and the reservoir properties 
to be oscillating with frequency $\omega = 2\pi/{\cal{T}}$. 

We use the scattering matrix approach to ac transport in 
phase coherent mesoscopic 
systems \cite{BTP94}. 
According to this approach the currents flowing in the system 
are determined by the scattering 
of electrons coming from the 
reservoirs by the mesoscopic sample
\cite{Buttiker90,Buttiker92}. 
In the present paper
we deal with non-interacting electrons.
A full theory has eventually to treat the internal potential 
of the pump in a self-consistent manner.

The scattering properties of a 
mesoscopic sample oscillating with frequency $\omega$
can be described via the Floquet
scattering matrix $\hat S^{\prime}_{F}$  (see, e.g.,
Ref. \onlinecite{MBstrong02}).

The matrix element $S^{\prime}_{F,\alpha\beta}(E_n,E)$ 
is a quantum mechanical amplitude 
for an electron with energy $E$ entering the scatterer through lead $\beta$ to 
leave the scatterer through lead $\alpha$ having 
energy $E_{n} = E + n\hbar\omega$. 
We use Greek letters $\alpha$, $\beta$ to number the leads
connecting the scatterer to the reservoirs: $\alpha, \beta = 1,\dots, N_{r}$. 

Denoting by $\hat a^{\prime}$ an annihilation operator for incoming particles 
we can write down the expression for the annihilation operators 
$\hat b^{\prime}$ for outgoing particles \cite{Buttiker92,MB02,MBstrong02}, 

\begin{equation}
\label{Eq2}
 \hat b^{\prime}_{\alpha}(E) = 
\sum_\beta \sum_{E_n>0} S^{\prime}_{F,\alpha\beta}(E,E_{n}) \hat a^{\prime}_{\beta}(E_n).
\end{equation}

By definition the reservoirs are not affected by the coupling to 
the scatterer
and thus they are in an equilibrium (but not necessary stationary) state.
Therefore the properties of incoming particles are independent of the 
scatterer and are determined by 
the reservoirs. To be definite we suppose that the cyclic evolution 
of any reservoir 
$\alpha$ is due to solely an oscillating electrochemical 
potential $\mu_{\alpha}(t)$:

\begin{equation}
\label{Eq3}
\begin{array}{l}
\mu_{\alpha}(t) = \mu_{0,\alpha} + eV_{\alpha}(t), \\
\ \\
V_{\alpha}(t) = V_{\alpha}\cos(\omega t + \varphi_\alpha), \\
\ \\
 eV_{\alpha}\ll \mu_{0,\alpha}.
\end{array}
\end{equation}

\noindent
We emphasize that we must keep track of the phase shifts $\varphi_{\alpha}$ 
since there are a number of oscillating potentials and we can not
eliminate all the phases $\varphi_{\alpha}$ 
simultaneously by merely shifting the time origin. 

It is well known (see e.g., Ref. \onlinecite{PB98}) 
that the wave functions 
for free electrons in the reservoir (say, $\alpha$) 
with an oscillating uniform potential 
$V_{\alpha}(t)$
are of the Floquet function type:

\begin{equation}
\label{Eq4}
 \psi_{\alpha}(E,t,{\bf r}) = e^{i{\bf kr} - i Et/\hbar}\sum\limits_{n=-\infty}^{\infty}
J_n\left(\frac{eV_{\alpha}}{\hbar\omega}\right)e^{-in(\omega t + \varphi_{\alpha})}.
\end{equation}

\noindent
Here $J_{n}(x)$ is the Bessel function of the first kind of 
the $n^{\rm th}$ order;
$E = \hbar^2k^2/(2m_e)$ ($m_e$ is an electron mass). 
The corresponding distribution function 
$f_{0,\alpha} = \langle\hat a^{\dagger}_{\alpha}(E)\hat a_{\alpha}(E)\rangle$
(here $\langle...\rangle$ means quantum-statistical averaging)
is independent of the oscillating potential $V_{\alpha}$ and 
is the Fermi distribution function

\begin{equation}
\label{Eq5}
f_{0,\alpha}(E) = \frac{1}{1 + \exp\left(\frac{E - \mu_{0,\alpha}}{k_BT_{\alpha}} \right)}.
\end{equation}

\noindent
Here $T_{\alpha}$ is the temperature of the reservoir $\alpha$; 
$k_B$ is the Boltzman constant.

In general, to find the Floquet scattering matrix $\hat S_F^{\prime}$,  
we have to investigate the transmission and reflection amplitudes 
of electrons with a wave function 
$\psi(E,t,{\bf r})$ given by Eq.(\ref{Eq4}) 
incident on the oscillating scatterer. 
However if the frequency $\omega$ is small compared with the energy of 
electrons participating in the transport (i.e., with the Fermi energy
$\mu$)

\begin{equation}
\label{Eq6}
\hbar\omega \ll \mu,
\end{equation}

\noindent
we can reduce the problem to scattering of ordinary plain waves.
To this end we use the following trick \cite{PB98}. 
We imagine that in the leads connecting scatterer to the 
reservoirs the oscillating potentials tend to zero: $V_{\alpha} = 0$. 
Then in the leads the electron wave functions are 
simply plain waves 

\begin{equation}
\label{Eq7}
\psi_{0,\alpha}(E,r) = e^{i{\bf kr} - i Et/\hbar}. 
\end{equation}

\noindent
In this region 
we introduce annihilation operators $\hat a$, $\hat b$ for 
incoming and outgoing particles, respectively. 
In close analogy with Eq.(\ref{Eq2}) they are related but
through the Floquet scattering matrix
$S_{F,\alpha\beta}(E_n,E)$ describing scattering of 
incident and outgoing plane waves:

\begin{equation}
\label{Eq8}
 \hat b_{\alpha}(E) = \sum_\beta \sum\limits_{n} 
S_{F,\alpha\beta}(E,E_n) \hat a_{\beta}(E_n).
\end{equation}

Comparing the wave functions Eq.(\ref{Eq4}) and Eq.(\ref{Eq7}) 
we see that the annihilation operators $\hat a$ for particles in the leads
can be expressed in terms of the annihilation operators $\hat a^\prime$ 
for particles in the reservoirs as follows \cite{PB98}

\begin{equation}
\label{Eq9}
 \hat a_{\alpha}(E) = \sum\limits_{n=-\infty}^{\infty}
J_n\left(\frac{eV_{\alpha}}{\hbar\omega}\right)e^{-in\varphi_{\alpha}}
 \hat a^\prime_{\alpha}(E - n\hbar\omega).
\end{equation}

\noindent
The above representation is valid for small frequencies Eq.(\ref{Eq6}).
Thus we
can put $k(E_n)\approx k(E)$ 
ignoring the terms of order $\hbar\omega/\mu$ and smaller.
In other words we ignore the reflection at the interface
between the region with oscillating potential and the region without one.

Using Eqs.(\ref{Eq8}) and (\ref{Eq9}) we calculate the distribution functions 
$f^{(out)}_{\alpha}(E) = \langle\hat b^{\dagger}_{\alpha}(E)\hat b_{\alpha}(E)\rangle$
for outgoing and 
$f^{(in)}_{\alpha}(E) =\langle\hat a^{\dagger}_{\alpha}(E)\hat a_{\alpha}(E)\rangle$
for incoming electrons in the leads as follows

\begin{subequations}
\label{Eq10}
\begin{equation}
\label{Eq10A} 
f^{(in)}_{\alpha}(E) = \sum\limits_{n=-\infty}^{\infty}
J_n^2\left(\frac{eV_{\alpha}}{\hbar\omega}\right) f_{0,\alpha}(E - n\hbar\omega),
\end{equation}
\ \\
\begin{equation}
\label{Eq10B} 
\begin{array}{l}
f^{(out)}_{\alpha}(E) = \sum\limits_{\beta}
\sum\limits_{n,m,q=-\infty}^{\infty} 
S^{*}_{F,\alpha\beta}(E,E_q) S_{F,\alpha\beta}(E,E_m) \\
\ \\
\times J_{n+q}\left(\frac{eV_{\beta}}{\hbar\omega}\right) 
J_{n+m}\left(\frac{eV_{\beta}}{\hbar\omega}\right) 
e^{i(q-m)\varphi_{\beta}} f_{0,\beta}(E - n\hbar\omega).
\end{array}
\end{equation}
\end{subequations}

Now the dc current $I_{\alpha}$ of spinless electrons, 
the quantity of interest here, 
flowing from the scatterer through the lead $\alpha$ can be expressed
in terms of these distributions \cite{MB02}, 

\begin{equation}
\label{Eq11}
 I_{\alpha} = \frac{e}{h} \int\limits_{0}^{\infty} dE 
\left\{f^{(out)}_{\alpha}(E) - f^{(in)}_{\alpha}(E)\right\}. 
\end{equation}

\noindent
Substituting Eqs.(\ref{Eq10}) into Eq.(\ref{Eq11}) we find

\begin{widetext}
\begin{equation}
\label{Eq12}
 I_{\alpha} = \frac{e}{h} \int\limits_{0}^{\infty} dE 
\sum\limits_{\beta} 
\sum\limits_{n=-\infty}^{\infty} 
f_{0,\beta}(E - n\hbar\omega) 
\left\{
\sum\limits_{m,q=-\infty}^{\infty} 
S^{*}_{F,\alpha\beta}(E,E_q) S_{F,\alpha\beta}(E,E_m) 
J_{n+q}\left(\frac{eV_{\beta}}{\hbar\omega}\right) 
J_{n+m}\left(\frac{eV_{\beta}}{\hbar\omega}\right)
e^{i(q-m)\varphi_{\beta}}
- \delta_{\alpha\beta} J_{n}^2\left(\frac{eV_{\alpha}}{\hbar\omega}\right) 
\right\}.
\end{equation}
\end{widetext}

Eq.(\ref{Eq12}) is the basic result which allows us to analyze
the dc currents flowing in the system under consideration. 
So far we put no restrictions on the reservoirs. 
Different temperatures of reservoirs as well as
different (stationary) electrochemical potentials can by themselves give 
rise to dc currents.  We will 
not consider the most general situation here.
Pumping in the presence of stationary chemical potential 
differences is investigated by Entin-Wohlman et al. \cite{EWAL02,EWAK03}.
Here we focus on dynamically oscillating potentials. 

In what follows we assume the reservoirs to have equal temperatures and 
equal dc-components of electrochemical potentials 
but the oscillating reservoir potentials $V_{\alpha}$ can be different:

\begin{equation}
\label{Eq13}
T_{\alpha} = T_{0},\quad \mu_{0,\alpha} = \mu_{0},\quad \alpha = 1,\dots , N_r. 
\end{equation} 

In this case the distribution functions entering Eq.(\ref{Eq12}) are 
independent of the lead index: $f_{0,\alpha(\beta)}(X) = f_{0}(X)$, 
where $f_{0}$ is the Fermi distribution
function with temperature $T_{0}$ and chemical potential $\mu_{0}$. 

To calculate the Floquet scattering matrix
$\hat S_{F}(E,E_n)$ one needs to solve the time-dependent scattering problem. 
Generally this can be done only numerically 
(see e.g., Ref.\onlinecite{MBstrong02}).

Here we are interested in the limit of low frequencies. 
In this limit we can use the adiabatic approximation
as a starting point and can express the Floquet scattering
matrix in terms of a stationary scattering matrix with time-dependent 
parameters (the {\it frozen} scattering matrix):
$\hat S_{0}(E,t) \equiv \hat S_{0}(E,\{P(t)\})$. 
Here $\{P\}$ is a set of parameters 
$P_{i}(t) = P_{i,0} + P_{i,1}\cos(\omega t + \phi_{i}), i=1,2,\dots, N_{p}$
oscillating with frequency $\omega$.   
The scattering matrix $\hat S_{0}(E,\{P\})$ describes
reflection and transmission of particles 
with energy $E$ at given frozen parameters $P_{i}$.
This description is valid if the energy scale $\hbar\omega$ dictated
by the modulation frequency is small compared with the energy scale
$\delta E$ over which the scattering matrix $\hat S(E)$ changes significantly
\cite{MBstrong02}.

\section{Adiabatic approximation}
\label{AA}

To zero-th order in frequency the elements of the Floquet scattering matrix 
can be approximated by the Fourier coefficients 
$\hat S_{0,n}$ of the stationary scattering matrix $\hat S_{0}$,
\begin{subequations}
\label{Eq14}
\begin{equation}
\label{Eq14A}
\hat S_{0,n}(E) = \frac{\omega}{2\pi}\int\limits_{0}^{\cal T} dt
e^{in\omega t} 
\hat S_{0}(E,t).
\end{equation}
\begin{equation}
\label{Eq14B}
\hat S_{0}(E,t) = \sum\limits_{n=-\infty}^{\infty}e^{-in\omega t} \hat
S_{0,n}(E), 
\end{equation}
\end{subequations}

\noindent
as follows \cite{MBstrong02}

\begin{equation}
\label{Eq15}
\hat S_{F}(E_n,E) \approx \hat S_{F}(E,E_{-n}) \approx \hat S_{0,n}(E). 
\end{equation}

\noindent
However in general this approximation is not sufficient to calculate
the current to order $\omega$.
In particular if the oscillating potentials $V_{\alpha}\neq 0$ are applied to the
reservoirs then to calculate the dc current to first order in frequency $\omega$
one needs to know the Floquet scattering matrix with the same accuracy.

Note that fortunately in the case of stationary reservoirs ($V_{\alpha} = 0$) 
there exists a representation (see Eq.(\ref{Eq8}) in Ref.\onlinecite{MBstrong02}) 
which allows to calculate the dc current (with accuracy of order $\omega$) 
using only the zero order approximation Eq.(\ref{Eq15}). 
In contrast another representation 
(see Eq.(\ref{Eq9}) in Ref.\onlinecite{MBstrong02}) 
for the same dc current requires the knowledge of the Floquet
scattering matrix with higher accuracy (i.e., to the first order in frequency).

Note that the nonadiabatic corrections to the scattering states and
the corresponding corrections
to the pumped current were considered in Refs.\onlinecite{EWAL02,EWAK03} 
in the limit of a small modulating potential. 
Our approach is valid for an arbitrary oscillating potential 
since we take into account the effect of all the harmonics of the pump
frequency $\omega$. 

To calculate the Floquet scattering matrix with an accuracy of order $\omega$ 
we generalize the approach used in Ref.\onlinecite{BTP94} 
and start from the unitarity
conditions for the Floquet scattering matrix \cite{MBstrong02}

\begin{subequations}
\label{Eq16}
\begin{equation}
\label{Eq16A}
\sum\limits_{\alpha}\sum\limits_{n=-\infty}^{\infty}
S^{*}_{F,\alpha\beta}(E_{n},E)S_{F,\alpha\gamma}(E_{n},E_{m}) 
= \delta_{m0}\delta_{\beta\gamma},
\end{equation}
\begin{equation}
\label{Eq16B}
\sum\limits_{\beta}\sum\limits_{n=-\infty}^{\infty}
S^{*}_{F,\alpha\beta}(E,E_{n})S_{F,\gamma\beta}(E_{m},E_{n}) 
= \delta_{m0}\delta_{\alpha\gamma}.
\end{equation}
\end{subequations}

\noindent
Taking into account that Eqs.(\ref{Eq15}) are a zero-th order approximation  
we will seek the first order approximation in the following form

\begin{subequations}
\label{Eq17}
\begin{equation}
\label{Eq17A}
\hat S_{F}(E_n,E) = \hat S_{0,n}\left(\frac{E_{n}+E}{2}\right) 
+ \hbar\omega\hat A_{n}(E) + O(\omega^2). 
\end{equation}
\begin{equation}
\label{Eq17B}
\hat S_{F}(E,E_{-n}) = \hat S_{0,n}\left(\frac{E+E_{-n}}{2}\right) 
+ \hbar\omega\hat A_{n}(E) + O(\omega^2). 
\end{equation}
\end{subequations}

\noindent
Here $\hat A_n(E)$ is a matrix of the Fourier coefficients for some matrix
$\hat A(E,t) \equiv \hat A(E,\{P(t)\})$ 
which is treated as independent of energy on the scale 
of the order of  $n\hbar\omega$;
$O(\omega^2)$ denotes the rest which is at 
least of second order in frequency
$\omega$ and which we neglect. 
Note that the first terms in Eqs.(\ref{Eq17}) should be expanded to 
the first order in $\omega$
$$
\hat S_{0,n}\left(\frac{E+E_{\pm n}}{2}\right) \approx \hat S_{0,n}(E) 
\pm \hbar\omega\frac{n}{2}\frac{\partial \hat S_{0,n}(E)}{\partial E}, 
$$ 
and other terms (of higher order in $\omega$) should be ignored.

Based on Eq.(\ref{Eq21}) we will show that Eq.(\ref{Eq17}) is, in fact,
an expansion in powers of $\hbar\omega/\delta E$.
Substituting Eqs.(\ref{Eq17}) into Eqs.(\ref{Eq16}) 
and keeping the terms of order $\omega^{0}$ and $\omega^{1}$
we get the required relations 
which can be used to calculate the current Eq.(\ref{Eq12}). 

In particular the diagonal part ($m=0, \beta=\gamma$) of Eqs.(\ref{Eq16}) gives 

\begin{equation}
\label{Eq18}
\begin{array}{c}
\sum\limits_{\alpha(\beta)}\sum\limits_{n=-\infty}^{\infty}
S^{*}_{0,\alpha\beta,n}(E)A_{\alpha\beta,n}(E) + c.c. =   \\
\ \\
\mp \frac{1}{2}\frac{\partial}{\partial E}
\sum\limits_{\alpha(\beta)}\sum\limits_{n=-\infty}^{\infty}
n|S_{0,\alpha\beta,n}(E)|^2.
\end{array}
\end{equation}

\noindent 
Here $c.c.$ denotes complex conjugate terms.
The sign $-(+)$ corresponds to the summation  over $\alpha (\beta)$. 

In what follows we concentrate on the case without magnetic fields and suppose that 
the stationary scattering matrix is symmetric in lead indexes:

\begin{equation}
\label{Eq19}
S_{0,\alpha\beta} =  S_{0,\beta\alpha}.  
\end{equation}

\noindent 
It follows from Eq.(\ref{Eq18}) that in this case 
the matrix $\hat A$ is antisymmetric:

\begin{equation}
\label{Eq20}
A_{\alpha\beta} =  - A_{\beta\alpha}.  
\end{equation}

\noindent
Since $A_{\alpha\alpha} = 0$, we can immediately conclude
that the reflection ($\alpha=\beta$) 
coefficients are with accuracy of order $\omega$
defined by the first terms on the RHS of Eqs.(\ref{Eq17}).
This fact justifies our representation 
for the elements of the Floquet scattering matrix in Eqs.(\ref{Eq17}).

We next need to determine the off-diagonal elements of $\hat A$.
The detailed calculation is given in Appendix \ref{A1} . 
The central result is the relation 
(valid to first order in $\omega$), 
\begin{subequations}
\label{Eq21}
\begin{equation}
\label{Eq21A}
\hbar\omega\left(\hat S_{0}^{\dagger}(E,t)\hat A(E,t) + \hat
A^{\dagger}(E,t)\hat S_{0}(E,t)\right)
= \frac{1}{2}{\cal P}\{\hat S_{0}^{\dagger};\hat S_{0} \}, 
\end{equation}
\begin{equation}
\label{Eq21B}
{\cal P}\{\hat S_{0}^{\dagger};\hat S_{0} \} = 
i\hbar \left( \frac{\partial \hat S_{0}^{\dagger}}{\partial t}
\frac{\partial \hat S_{0}}{\partial E} -
\frac{\partial \hat S_{0}^{\dagger}}{\partial E}
\frac{\partial \hat S_{0}}{\partial t}
\right).
\end{equation}
\end{subequations}

\noindent
Here $i$ is the imaginary unit. 
Since the scattering matrix is unitary $\hat S_0^{\dagger}\hat S_0 = \hat I$ 
(where $\hat I$ is a unit matrix) the matrix
${\cal P}\{\hat S_{0}^{\dagger};\hat S_{0} \}$ is traceless:
$\sum\limits_{\alpha=1}^{N_r}{\cal P}\{\hat S_{0}^{\dagger};\hat S_{0} \}_{\alpha\alpha}= 0$.
(see Appendix \ref{A2} for the detailed proof).

Note Avron et al.\cite{AEGS03} consider a closely related matrix 
$\hat \Omega = {\cal P}\{\hat S_{0};\hat S_{0}^{\dagger}\}$.
This matrix is the commutator 
$\hat \Omega = \frac{i}{\hbar}[\hat {\cal T},\hat {\cal E}]$
of the Wigner time-delay matrix \cite{Wigner55,Smith60}:
$\hat{\cal T} = -i\hbar\frac{\partial \hat S_0}{\partial E}\hat S_{0}^{\dagger}$,
and the matrix of the energy shift \cite{AEGS01,AEGS03}:
$\hat{\cal E} = i\hbar\frac{\partial \hat S_0}{\partial t}\hat S_{0}^{\dagger}$.
Note however, that on the RHS of Eq. (\ref{Eq21A}) the
commutator appears with a different sequence of $S^{\dagger}$ and $S$ as 
compared to $\Omega$. For this reason (and other reasons to be become 
clear later on, we have introduced a separate notation, the Poisson 
bracket ${\cal P}$. As we will show (see Eq.(\ref{Eq34}))  
the diagonal elements 
$ (e/h) {\cal P}_{\alpha\alpha}$ of this commutator
are just spectral current densities (current per energy). 
 
From Eq.(\ref{Eq21A}) we find an expression for the product 
of the frozen scattering matrix with elements of $\hat A$, 

\begin{equation}
\label{Eq21+}
4\hbar\omega Re[S^{*}_{0,\alpha\beta}A_{\alpha\beta}] = 
\frac{1}{N_r}\bigg(
{\cal P}\{\hat S_{0}^{\dagger};\hat S_{0} \}_{\beta\beta} 
- {\cal P}\{\hat S_{0}^{\dagger};\hat S_{0} \}_{\alpha\alpha} \bigg).
\end{equation}

\noindent
Note that the scattering matrix $\hat S_{0}$ is a unitary matrix 
of dimension $N_r$.
Evidently Eq.(\ref{Eq21+}) is consistent with Eq.(\ref{Eq20}).

Below we use Eqs.(\ref{Eq17}) and (\ref{Eq21+}) 
to evaluate the current Eq.(\ref{Eq12}) with an accuracy of order $\omega$.

\section{Linear response adiabatic current}
\label{LR}

Now we use the adiabatic approximation introduced in the previous section 
and calculate the 
zero-frequency, dc-current Eq.(\ref{Eq12}) to linear order in the oscillating 
potentials $V_{\alpha}\to 0$ of the reservoirs at finite temperature $T_{0}$. 
We assume that the following conditions hold:

\begin{subequations}
\label{Eq22}
\begin{equation}
\label{Eq22A}
\hbar\omega \ll k_BT_{0},
\end{equation}
\begin{equation}
\label{Eq22B}
eV_{\alpha} \ll k_BT_{0}.
\end{equation}
\end{subequations}

The first inequality ($\hbar\omega\ll k_BT_{0}$) is relevant for experiments 
on adiabatic ($\omega\to 0$) quantum transport. 
The second inequality defines nothing but the
linear response regime.

In Eq.(\ref{Eq12}) the sum over $n$ contains 
approximately $n_{max}\sim\frac{eV}{\hbar\omega}$ terms. 
Therefore, $\hbar\omega n \leq eV$ and because of Eq.(\ref{Eq22B})
we have $\hbar\omega n \ll k_BT$.
Hence we can expand the Fermi function entering Eq.(\ref{Eq12}). 
Taking into account Eq.(\ref{Eq13}) this expansion (up to second 
order in $\omega$) is:
$f_{0,\beta}(E - n\hbar\omega) \approx f_{0}(E) - 
n\hbar\omega\frac{\partial f_{0}(E)}{\partial E}
+ \frac{1}{2}n^2\hbar^2\omega^2\frac{\partial^2 f_{0}(E)}{\partial E^2}$.
Substituting this distribution  into Eq.(\ref{Eq12}) and take the sum over $n$. 
We use the summation formulae for the Bessel functions \cite{Bateman}

\begin{equation}
\label{Eq23}
\begin{array}{l}
\sum\limits_{n=-\infty}^{\infty}J_{n+m}(X)J_{n+q}(X) = \delta_{mq}, \\
\ \\
\sum\limits_{n=-\infty}^{\infty}nJ_{n+m}(X)J_{n+q}(X) = 
-m\delta_{mq} + \frac{X}{2}(\delta_{m(q+1)} + \delta_{m(q-1)}), \\
\ \\
\sum\limits_{n=-\infty}^{\infty}n^2J_{n+m}(X)J_{n+q}(X) = 
\left(m^2 + \frac{X^2}{2}\right)\delta_{mq} \\
\ \\
\begin{array}{r}
- X\left( [m-0.5]\delta_{m(q+1)} + [m+0.5]\delta_{m(q-1)} \right) \\
\ \\
+ \frac{X^2}{4}(\delta_{m(q+2)} + \delta_{m(q-2)}).
\end{array}
\end{array}
\end{equation}

After that, substituting Eqs.(\ref{Eq17}), (\ref{Eq21+})
and applying the inverse Fourier transformation Eq.(\ref{Eq14B}), 
we sum over $q$ and $m$.
Finally we represent the dc current $I_{\alpha}$ flowing in 
lead $\alpha$ under the action 
of an oscillating scatterer and oscillating reservoir potentials 
in the following way:

\begin{widetext}
\begin{subequations}
\label{Eq24}
\begin{equation}
\label{Eq24A}
I_{\alpha} =   \int\limits_{0}^{\infty} dE 
\left(-\frac{\partial f_0(E)}{\partial E} \right)  
\left\{ I_{\alpha}^{(pump)} + I_{\alpha}^{(rect)} + I_{\alpha}^{(int)} \right\}, 
\end{equation}
\ \\
\begin{equation}
\label{Eq24B}
I_{\alpha}^{(pump)}(E) =  i\frac{e}{2\pi}
\overline{
\left( \frac{\partial\hat S_{0}(E,t)}{\partial t} \hat S_{0}^{\dagger}(E,t) \right)_{\alpha\alpha}
}, 
\end{equation}
\ \\
\begin{equation}
\label{Eq24C}
I_{\alpha}^{(rect)}(E) =  G_0 \sum\limits_{\beta} 
\overline{ \big(V_{\beta}(t) - V_{\alpha}(t) \big)  \big|S_{0,\alpha\beta}(E,t)\big|^2 },
\end{equation}
\ \\
\begin{equation}
\label{Eq24D}
I_{\alpha}^{(int)}(E) =  \frac{G_0}{2N_r}\sum\limits_{\beta} 
\overline{V_{\beta}(t)
\left( 
{\cal P}\{\hat S_{0}^{\dagger};\hat S_{0} \}_{\beta\beta} 
- {\cal P}\{\hat S_{0}^{\dagger};\hat S_{0} \}_{\alpha\alpha} 
- N_r{\cal P}\{S^{*}_{0,\alpha\beta}; S_{0,\alpha\beta}\} \right)
}. 
\end{equation}
\end{subequations}
\end{widetext}

\noindent
Here the bar denotes the time average 
$\overline{X} = \frac{1}{\cal T}\int\limits_{0}^{\cal{T}}dt X(t)$
over a time period ${\cal T} = \frac{2\pi}{\omega}$;
$G_0 = e^2/h$ is the spinless conductance quantum;
the function ${\cal P}\{X;Y\}$ is defined in Eq.(\ref{Eq21B}).
To arrive at Eq.(\ref{Eq24C}) we used the unitarity condition 
$\sum\limits_{\beta}\big| S_{0,\alpha\beta}\big|^2 = 1$ and 
the fact that the average potential is zero:
$\overline{V_{\alpha}(t)} = 0$.

We emphasize that in the above expressions we omitted all the terms which are
of the second (and higher) order in frequency $\omega$ and/or in 
potentials $V_{\alpha}$.
Next we characterize briefly the three contributions to the 
current $I_{\alpha}$.

The current $I_{\alpha}^{(pump)}$ is due to solely the oscillating scatterer.
It determines the quantum pump effect when the reservoirs are stationary. 
It is the formula obtained by Brouwer \cite{Brouwer98}.

The current $I_{\alpha}^{(rect)}$ is a 
consequence of the rectification of ac currents flowing in the
system under the influence of ac potentials $V_{\alpha}$ 
applied to the reservoirs. 
In context of pumping this effect was considered 
by Brouwer in Ref. \onlinecite{Brouwer01}. 
Note that this rectified current depends on the conductances
$G_{\alpha\beta} = - G_0|S_{\alpha\beta}|^2$ and the 
corresponding potential differences
$\Delta V_{\alpha\beta}(t) = V_{\beta}(t) - V_{\alpha}(t)$ 
in close analogy with the dc current flowing in response to a dc voltage. 
We stress that here $\Delta V_{\alpha\beta}(t)$ 
depends not only on the amplitudes of the corresponding potentials but
also on the phase lag 
$\Delta\varphi_{\alpha\beta} = \varphi_{\beta} - \varphi_{\alpha}$ as well.
In particular if the amplitudes of two oscillating potentials are equal
$V_{\alpha} = V_{\beta} = V_{0}$ then the potential difference reads
 
\begin{equation}
\label{Eq25}
 \Delta V_{\alpha\beta}(t) = -2V_{0}\sin(\Delta\varphi_{\alpha\beta})
\sin\left(\omega t + \frac{\varphi_{\alpha} + \varphi_{\beta}}{2} \right).
\end{equation}

\noindent 
This equation [together with Eq.(\ref{Eq24C})] shows clearly that 
the rectification of ac currents can depend on the phase lag between the 
applied ac potentials and, hence, it can mimic a quantum pump 
effect \cite{Brouwer01}.

The third term $I_{\alpha}^{(int)}$ is novel. Interestingly, 
as we will see, this current renormalizes both  
$I_{\alpha}^{(pump)}$ and $I_{\alpha}^{(rect)}$.
The current $I_{\alpha}^{(int)}$ is a consequence  
of the {\it interference} between 
the ac currents produced by the external voltages 
and the ac currents produced by the nonstationary scatterer.
Remarkably, it is essentially determined by commutator expressions. 
An "oscillating" scatterer is much richer in physics then
expressed by Eq. (\ref{Eq24C}). 
The expression for $I^{(rect)}$ is widely used but this is only a part of a correct answer. 
The part $(I^{(rect)})$ is due to a rectification of external currents caused by 
the time-dependence of the conductances. 
The oscillating scatterer is much richer: It generates its own ac currents which can
interfere with the external ac currents. 
This interference effect leads to $I^{(int)}$.

Before proceeding we check the current conservation. 
To this end we sum  $I_{\alpha}$ over the lead index $\alpha$. 
Note that each of the currents
$I_{\alpha}^{(pump)}$, $I_{\alpha}^{(rect)}$, and $I_{\alpha}^{(int)}$
is separately conserved. 
This fact supports the current decomposition introduced above. 

For the pump currents $I_{\alpha}^{(pump)}$, 
using the Birman-Krein formula \cite{Yafaev92} we find
$$
\sum\limits_{\alpha}I_{\alpha}^{(pump)} 
\sim Tr \left(
\overline{  \frac{\partial\hat S_{0}}{\partial t} \hat S_{0}^{\dagger} } \right) 
= \overline{  \frac{\partial}{\partial t}\ln(\det\hat S_{0}) }  = 0.
$$
Here we take into account that the average of a time derivative is identically zero:
$\overline{\frac{\partial X(t)}{\partial t}}\equiv 0$;
$Tr$ denotes the trace of a matrix: 
$Tr \hat S = \sum\limits_{\alpha} S_{\alpha\alpha}$.

The conservation of the rectification currents 
$\sum\limits_{\alpha}I_{\alpha}^{(rect)} =
G_0\sum\limits_{\alpha,\beta} \overline{
\big( V_{\beta}(t) - V_{\alpha}(t) \big) \big|S_{0,\alpha\beta}\big|^2} = 0$
follows from the unitarity condition
$\sum\limits_{\alpha} \big|S_{0,\alpha\beta}\big|^2 = 
\sum\limits_{\beta} \big|S_{0,\alpha\beta}\big|^2 = 
1$.

The current $I_{\alpha}^{(int)}$ is conserved as well. 
Since the matrix ${\cal P}\{\hat S^{\dagger}_{0}; \hat S_{0}\}$ Eq.(\ref{Eq21B})
is traceless we get from Eq.(\ref{Eq24D})

\begin{widetext}
$$
\sum\limits_{\alpha=1}^{N_r}I_{\alpha}^{(int)} = \frac{G_0}{2N_r}\sum\limits_{\beta} 
\overline{ V_{\beta}(t) \sum\limits_{\alpha=1}^{N_r}\left(
{\cal P}\{\hat S^{\dagger}_{0}; \hat S_{0}\}_{\beta\beta}
- {\cal P}\{\hat S^{\dagger}_{0}; \hat S_{0}\}_{\alpha\alpha}
- N_r {\cal P}\{S^{*}_{0,\alpha\beta}; S_{0,\alpha\beta}\}
\right) } \sim N_r{\cal P}\{\hat S^{\dagger}_{0}; \hat S_{0}\}_{\beta\beta}
- N_r{\cal P}\{\hat S^{\dagger}_{0}; \hat S_{0}\}_{\beta\beta} = 0.
$$
\end{widetext}

To shed more insight onto the nature of the new contribution $I^{(int)}$ 
we consider a simple but a quite generic example.

\subsection{Two terminal single channel scatterer}
\label{TT}

Consider a nonstationary scatterer connected to only two reservoirs 
$\alpha = 1,2$ via single channel leads. 
For such a scatterer, assuming there are no magnetic fields,  
the stationary scattering matrix $\hat S_{0}$ 
is a symmetric $2\times 2$ unitary matrix.

\begin{equation}
\label{Eq26}
\hat S_{0} = e^{i\gamma}\left(
  \begin{array}{cc}
       \sqrt{R}e^{-i\theta}   & i\sqrt{T}            \\
       i\sqrt{T}             & \sqrt{R}e^{i\theta} \\
  \end{array}
\right).
\end{equation}

\noindent
Here $R$ and $T$ are the reflection and the transmission
probability, respectively ($R+T=1$). 
The phase $\theta$ characterizes the asymmetry between 
the reflection to the left and to the right. 
The phase $\gamma$ relates to the change of the overall charge $\delta Q$
on the scatterer (for instance a dot) via the Friedel 
sum rule \cite{Friedel52}:
$\delta\gamma= \pi\delta Q/e$ (where $e$ is the electron charge), 
or in different notation
$ \delta Q = e/(2\pi i)\delta[\ln\det\hat S] $.
We assume that $R,T = 1- R ,\theta,\gamma$ are functions 
of the electron energy $E$ and the external parameters 
$P_i(t)$ varying with frequency $\omega$. 
Before proceeding we remark that 
for the case $N_r = 2$ the current $I_{\alpha}^{(int)}$ 
Eq. (\ref{Eq24D}) can be simplified

\begin{equation}
\label{Eq27}
\begin{array}{c}
I_{\alpha}^{(int)} =  \frac{G_0}{2}\left(
\overline{V_{\beta}(t){\cal P}\{S^{*}_{0,\beta\beta}; S_{0,\beta\beta}\} } \right. \\
\ \\
\left.
- \overline{V_{\alpha}(t) {\cal P}\{S^{*}_{0,\alpha\alpha}; S_{0,\alpha\alpha}\} } \right),
\quad \alpha\neq\beta.
\end{array}
\end{equation}

\noindent
Substituting the scattering matrix Eq.(\ref{Eq26}) into Eqs.(\ref{Eq24})
and (\ref{Eq27}) we find the currents $I_{1} = - I_{2}$
flowing between the scatterer and the reservoirs:

\begin{subequations}
\label{Eq28}
\begin{equation}
\label{Eq28A}
I_{1}^{(pump)}(E) = \frac{e}{2\pi}
\overline{ R(E,t) \frac{\partial \theta(E,t)}{\partial t} }, 
\end{equation}
\begin{equation}
\label{Eq28B}
I_{1}^{(rect)}(E) =  G_0
\overline{ T(E,t)[V_{2}(t) - V_{1}(t)]},
\end{equation}
\begin{equation}
\label{Eq28C}
\begin{array}{c}
I_{1}^{(int)}(E) =  
\frac{e^2}{4\pi} \overline{ 
\left( \frac{\partial \theta}{\partial t}\frac{\partial R}{\partial E} - 
        \frac{\partial \theta}{\partial E}\frac{\partial R}{\partial t} \right)
[V_{2}(t) + V_{1}(t)] } \\
\ \\
\quad\quad\quad + \quad  
\frac{e^2}{4\pi} \overline{ 
\left( \frac{\partial \gamma}{\partial t}\frac{\partial R}{\partial E} - 
        \frac{\partial \gamma}{\partial E}\frac{\partial R}{\partial t} \right)
[V_{2}(t) - V_{1}(t)] }. 
\end{array}
\end{equation}
\end{subequations}
These expressions demonstrate 
that the current $I^{(int)}$ has 
common features with both the rectification current $I^{(rect)}$ 
and the pumped current $I^{(pump)}$. 
Like the former, the current  $I^{(int)}$ depends on the potential difference
$\Delta V_{12}$. 
Like the latter, the current $I^{(int)}$ can exist
even at equal reservoir potentials $V_{1}(t) = V_{2}(t)$.
In this case, the conditions necessary for the existence of 
$I^{(int)}$ and $I^{(pump)}$
are the same \cite{MBstrong02}: First, the scatterer has to be asymmetric, 
i.e., $\theta\neq 0$,
and, second, the time reversal symmetry (TRS) has to be broken. 
We note that the current $I^{(int)}$ depends on both
the oscillating reservoir potentials 
$V_{\alpha}(t) = V_{\alpha}\cos(\omega t + \varphi_{\alpha})$ 
and the oscillating scattering parameters
$P_{i}(t) = P_{i,0} + P_{i,1}\cos(\omega t + \phi_i)$.
Therefore analyzing the presence/absence of the TRS we have to consider
all the phases, namely $\varphi_{\alpha}$ as well as $\phi_i$.

We have here treated only non-interacting electrons. 
As a consequence sums of potentials appear in Eq.(\ref{Eq28C}).
This is in contrast with an electrically self-consistent theory 
which permits only the appearance of voltage differences.
If interactions are switched on \cite{BTP94} then the (self-consistent)
potential  $U\neq 0$ 
inside the scatterer becomes dependent on external potentials
$V_{\alpha}$ and the 
differences $V_{\alpha} - U$ should appear instead of $V_{\alpha}$.
$U$ is in general a function of all the oscillating parameters $P_{i}(t)$,
all the external potentials $V_{\alpha}$ and also of the potentials 
at the gates which influence the electrostatic potential inside the 
scatterer. 
Our expressions do, however, conserve 
current.

We see that the first term on the RHS of Eq.(\ref{Eq28C})
renormalizes the pumped current $I^{(pump)}$ and 
the second one renormalizes the rectification current $I^{(rect)}$.
The latter is due to nonadiabatic (first order in $\omega$) corrections to
the conductances
arising from the corresponding corrections Eq.(\ref{Eq17}) to the
scattering matrix.
Note that the analogous corrections are discussed in
Refs.\onlinecite{EWAL02,EWAK03}
in context of pumping in the presence of a dc bias.

Since the pump effect is the main topic of this work we consider
now the case with $V_1 (t) = V_2 (t) $ in more detail.
This case corresponds to an experimental setup in which 
the scatterer and a large
portion of the reservoirs to which it is 
connected are subject to long wavelength radiation. 
The effect of such radiation can be modeled via an  oscillating uniform 
potential $V(t)$
which is the same at different reservoirs: $V_1(t) = V_2(t)\equiv V(t)$.
In this case the rectification current Eq.(\ref{Eq28B}) is 
absent $I^{(rect)} = 0$, 
and the whole dc current $I_{\alpha}$ can be reduced to the simple form

\begin{equation}
\label{Eq29}
\begin{array}{c}
I_{\alpha} =  \frac{e}{2\pi} \int\limits_{0}^{\infty} dE 
\left(-\frac{\partial f_0(E)}{\partial E} \right)   
\overline{ R({\cal E},t) \frac{d \theta({\cal E},t)}{d t} }, \\
\ \\
{\cal E} = E + eV(t).
\end{array}
\end{equation}

\noindent 
To obtain this result we have used the following identity: 
$\overline{-A\frac{\partial R}{\partial t}} = 
\overline{R\frac{\partial A}{\partial t}}$
with $A = eV\frac{\partial\theta}{\partial E}$.
We have also introduced the full time derivative:
$\frac{d}{dt} = \frac{\partial}{\partial t} + 
e\frac{dV}{dt}\frac{\partial}{\partial {\cal E}}$.

This result can be understood in the following way: 
For stationary reservoirs ($V(t)=0$)
the pumped current is described by equations (\ref{Eq24A}) and (\ref{Eq28A}) 
with the quantities $R$ and $\theta$ taken at the energy $E$ of 
incident electrons. However, 
if the chemical potential $\mu(t) = \mu_{0} + eV(t),~V(t)\neq 0$ 
oscillates slowly ($\omega\to 0$) 
then we can consider incident electrons having energy 
${\cal E} = E + eV(t)$ following
adiabatically the reservoir's potential $V(t)$. 
Substituting in Eq.(\ref{Eq28A}) ${\cal E}$ instead of $E$ 
and replacing a partial time derivative by a full time 
derivative we get Eq.(\ref{Eq29}). 

It should be noted that the above substitution ${\cal E} = E + eV(t)$
implies that the potential inside the scatterer ($U=0$) is 
independent of the external potentials $V_{\alpha}$. This is correct
for noninteracting electrons but it should be modified if the 
interactions are present
\cite{BTP94}.

From equation (\ref{Eq29}) we can conclude that the effect of an 
oscillating external 
potential $V(t)$ is like the effect generated by 
an oscillating parameter of the scatterer 
(i.e., an oscillating internal potential). 
Therefore to analyze the ability of an open system  
(the scatterer plus reservoirs)
to generate adiabatic dc currents we have 
to consider the full set of oscillating parameters
$\{V_{\alpha}(t), P_{i}(t)\}, (\alpha=1,2,\dots,N_{r}; i = 1,2,\dots,N_p)$.

\section{dc current at finite ac voltages}
\label{FV}

Now we go beyond linear response theory. 
We suppose that the potentials $V_{\alpha}$ 
can be large compared to the temperature.
Thus we calculate the current Eq.(\ref{Eq12}) with accuracy 
up to the first order in
$\omega$ and with an arbitrary ratio
of the potentials $V_{\alpha}$ to the temperature: 

\begin{subequations}
\label{Eq31}
\begin{equation}
\label{Eq31A}
\hbar\omega \ll k_BT_{0},
\end{equation}
\begin{equation}
\label{Eq31B}
eV_{\alpha} \ll \mu_{0,\alpha}.
\end{equation}
\end{subequations}

Since the potentials $V_{\alpha}$  are not necessarily small
compared with the temperature $T$ we can not expand 
the Fermi function $f_{0,\beta}(E-n\hbar\omega)$ entering Eq.(\ref{Eq12}). 
Nevertheless Eq.(\ref{Eq31A}) allows us to sum over $n$ 
and to simplify Eq.(\ref{Eq12}). 

To this end we go from the energy representation over to the time 
representation.
We express the Fermi function $f_{0,\beta}(E)$ Eq.(\ref{Eq5})
and the Bessel functions $J_n(x)$ as follows:

$$
f_{0,\beta}(E-n\hbar\omega) = 
\int\limits_{-\infty}^{\infty} d\tau f_{0,\beta}(\tau)
e^{i(E-n\hbar\omega)\frac{\tau}{\hbar}}, 
$$
$$
J_{n+q}\left(\frac{eV_{\beta}}{\hbar\omega}\right)e^{i\varphi_{\beta}(n+q)} =
\frac{1}{\cal T}\int\limits_{0}^{\cal T} dt W^{*}_{\beta}(t) e^{-i(n+q)\omega t}, 
$$
$$
J_{n+m}\left(\frac{eV_{\beta}}{\hbar\omega}\right)e^{-i\varphi_{\beta}(n+m)} =
\frac{1}{\cal T}\int\limits_{0}^{\cal T} dt_{1} W_{\beta}(t_{1}) e^{i(n+m)\omega t_{1}}, \\
$$
$$
f_{0,\beta}(\tau) = \frac{ik_BT_{0}}{2\hbar\sinh\left(\pi k_BT_{0}\frac{\tau}{\hbar} \right)}
e^{-i\mu_{0,\beta}\frac{\tau}{\hbar}}, 
$$
$$
W_{\beta}(t) = e^{-i\frac{eV_{\beta}}{\hbar\omega}\sin(\omega t + \varphi_{\beta})}.
$$

Substituting these equations into Eq.(\ref{Eq12}) and summing over $n$ we obtain
a delta-function $\delta(t_1 - t - \tau)$ which allows us to perform one additional integration.
At $\tau > 0$ ($\tau < 0$) we integrate over $t_{1}$ ($t$). This leads 
to the substitution $t_1 = t + \tau$ ($t = t_1 + |\tau|$). 
Further we expand $\sin(\omega \tau + \omega t + \varphi_{\beta})$ to 
first order
in $\omega\tau$. We can do this because for the relevant 
$\tau \leq \frac{\hbar}{k_BT_{0}}$
Eq. (\ref{Eq31A}) gives $\omega\tau\ll 1$.
Next we integrate over $\tau$ and finally get the dc current as follows:

\begin{widetext}
\begin{equation}
\label{Eq32}
 I_{\alpha} = \frac{e}{h} \int\limits_{0}^{\infty} dE 
\frac{1}{\cal T}\int\limits_{0}^{\cal T} dt
\left\{
 \sum\limits_{\beta}\sum\limits_{m,q=-\infty}^{\infty} 
f_{0}\Big(E + (q+m)\frac{\hbar\omega}{2};\mu_{\beta}(t)\Big) 
S^{*}_{F,\alpha\beta}(E,E_q) S_{F,\alpha\beta}(E,E_m) 
e^{i(m-q)\omega t}
- f_{0}\big(E;\mu_{\alpha}(t)\big) 
\right\}.
\end{equation}
\end{widetext}

\noindent
Here we have introduced the Fermi function with time-dependent chemical potential
$\mu_{\alpha}(t) = \mu_{0,\alpha} + eV_{\alpha}(t)$ Eq.(\ref{Eq3}):
$f_{0}\big(E;\mu_{\alpha}(t)\big)  = 
\left[1 + \exp\left(\frac{E-\mu_{\alpha}(t)}{k_BT_{0}} \right) \right]^{-1}$.

Note that Eq.(\ref{Eq32}) is valid both for the adiabatic 
as well as for the nonadiabatic case. 
The only restriction is that the frequency has to be small compared with the temperature 
Eq.(\ref{Eq31A}).

Next we use the adiabatic approximation of Sec.\ref{AA} and calculate 
the current $I_{\alpha}$ to first order in frequency $\omega$ 
under the conditions of Eq.(\ref{Eq13}).
To this end we substitute Eqs.(\ref{Eq17}) and (\ref{Eq21+})  into 
Eq.(\ref{Eq32}) 
and expand the Fermi function in powers of $\omega$.
Next we use the inverse Fourier transformation  Eq.(\ref{Eq14B})
and after a little manipulation 
(we integrate by parts on energy and dropped the 
contribution arising from $E=0$; 
in addition we exploit the unitarity of the frozen scattering matrix
$\sum\limits_{\alpha}\big| S_{0,\alpha\beta}(E,t)\big|^2=1$) and 
find the current:

\begin{widetext}
\begin{equation}
\label{Eq33}
I_{\alpha} =   \frac{e}{h}
\int\limits_{0}^{\infty} dE \frac{1}{\cal T}\int\limits_{0}^{\cal T} dt
\left\{ 
\sum\limits_{\beta} f_{0}\big(E;\mu_{\beta}(t)\big)
\left[ \big|S_{0,\alpha\beta}(E,t)\big|^2 
+ \frac{ {\cal P}\{\hat S_{0}^{\dagger};\hat S_{0} \}_{\beta\beta} 
- {\cal P}\{\hat S_{0}^{\dagger};\hat S_{0} \}_{\alpha\alpha} 
- N_r{\cal P}\{S^{*}_{0,\alpha\beta}; S_{0,\alpha\beta}\} }{2N_r} \right]
 - f_{0}\big(E;\mu_{\alpha}(t)\big)
\right\} . 
\end{equation}
\end{widetext}

\noindent 
The above equation generalizes Eqs.(\ref{Eq24}) to the case of finite voltages. 
Current conservation $\sum\limits_{\alpha} I_{\alpha} = 0$ can easily be proven
in analogy with Eqs.(\ref{Eq24}).

Next we concentrate on the pump effect and
consider the case with reservoirs having equal oscillating potentials:
$\mu_{\alpha}(t) \equiv\mu(t)=  \mu_{0} + eV\cos(\omega t + \varphi),~~ \alpha = 1,\dots,N_r$.
Since the Fermi functions entering Eq.(\ref{Eq33}) become independent of the lead index 
we can sum up over $\beta$  and obtain

\begin{equation}
\label{Eq34}
\begin{array}{l}
I_{\alpha} =  
\int\limits_{0}^{\infty} dE \frac{1}{\cal T}\int\limits_{0}^{\cal T} dt
f_{0}\big( E;\mu(t)\big) \frac{dI_{\alpha}(E,t)}{dE}, \\
\ \\
\frac{dI_{\alpha}}{dE} = 
 \frac{e}{h}{\cal P}\{\hat S_{0};\hat S^{\dagger}_{0}\}_{\alpha\alpha}
\equiv i\frac{e}{2\pi} 
\left(  \frac{\partial\hat S_{0}}{\partial t}\frac{\partial\hat S_{0}^{\dagger}}{\partial E} - 
\frac{\partial\hat S_{0}}{\partial E}\frac{\partial\hat S_{0}^{\dagger}}{\partial t}
\right)_{\alpha\alpha},
\end{array}
\end{equation}

\noindent
Here we used an obvious equality:
${\cal P}\{\hat S^{\dagger}_{0};\hat S_{0}\}_{\alpha\alpha} =
-{\cal P}\{\hat S_{0};\hat S^{\dagger}_{0}\}_{\alpha\alpha} $.

The quantity $dI_{\alpha}(E,t)/dE$
is the spectral current density 
at energy $E$ and time $t$,  
(i.e., the current within the energy interval $dE$) 
produced by the adiabatically evolving scatterer towards the reservoir $\alpha$.
This definition seems reasonable because of a conservation law 
$\sum\limits_{\alpha}dI_{\alpha}/dE(E,t) = 0$ 
which is valid at any energy $E$ and at any time moment $t$. 
Note that in the case of stationary reservoirs the same interpretation was given
in Ref. \onlinecite{AEGS03}.

These currents (or more precisely, the ability to produce them) 
are an intrinsic property of a time-dependent scatterer. 
This property differentiates between a nonstationary scatterer and a "frozen" one.
Note that the Fermi distribution function in Eq.(\ref{Eq34}) describe 
the filling of (potentially) existing "current" states of a nonstationary scatterer.

At $V=0$ the Eq.(\ref{Eq34}) reproduces Brouwer's result Eq.(\ref{Eq24B})
and agrees with that obtained in Ref.\onlinecite{AEGS03}.
At small voltages $V\to 0$ for the scattering matrix Eq.(\ref{Eq26}) 
we get Eq.(\ref{Eq29}).

Equation (\ref{Eq34}) determines the dc-current to the first order in $\omega$
pumped by the slowly oscillating scatterer
between the reservoirs having equal (possibly zero)
oscillating potentials $V_{\alpha}(t) = V(t)$.
Formally in the adiabatic case under consideration 
the effect of oscillating chemical potentials is only 
the change of an energy of electrons falling upon the scatterer. 
However, in fact, the phase $\varphi$ of an oscillating potential 
$V(t) = V\cos(\omega t +\varphi)$ is of a great importance 
because of the following. 
An adiabatically pumped current $I_{\alpha}\neq 0$ is generated 
already if the time reversal symmetry is broken in the {\it whole system} 
including the scatterer and the reservoirs. 
At $V\neq 0$ analyzing this question we have to take into account 
a possible phase shift between the potentials of reservoirs 
and the oscillating parameters 
$P_i(t) = P_{i,0} + P_{i,1}\cos(\omega t + \phi_i)$ of a scatterer. 
In particular even a scatterer with a {\it single} oscillating parameter 
can produce an adiabatic dc current if only $\phi_1\neq\varphi$.

\section{Instantaneous current}
\label{IC}

In this Section we derive an expression for the instantaneous current
of an adiabatic quantum pump simultaneously subject to oscillating 
external potentials. 
We first clarify the physical 
meaning of the (diagonal elements of the) 
quantity ${\cal P}\{\hat S_{0}^{\dagger};\hat S_{0}\}$
defining (antisymmetric in lead indices) nonadiabatic 
corrections Eqs.(\ref{Eq21})
to the scattering matrix. 
From the geometrical point of view \cite{AEGS03}
this quantity is a curvature in the time-energy plane.
The physical interpretation is based on Eq.(\ref{Eq34}). 
We can consider the quantity
$dI_{\alpha}/dE(E,t) = \frac{e}{h}{\cal P}\{\hat S_{0};\hat S^{\dagger}_{0}\}_{\alpha\alpha}$ 
as an instantaneous {\it spectral current} which is 
pushed by the oscillating scatterer into the lead $\alpha$.
A more detailed partitioning of the current follows from Eq.(\ref{Eq33}):
We can say that the scatterer drives 
the following spectral currents
from the lead $\beta$ into the lead $\alpha$:
\begin{equation}
\label{Eq34_A}
\frac{dI_{\alpha\beta}}{dE} = \frac{e}{h}\frac{
  {\cal P}\{ \hat S_{0}; \hat S_{0}^{\dagger} \}_{\alpha\alpha} 
- {\cal P}\{ \hat S_{0}; \hat S_{0}^{\dagger} \}_{\beta\beta} 
+ N_r{\cal P}\{S_{0,\alpha\beta}; S^{*}_{0,\alpha\beta}\}
}{2N_r}.
\end{equation}
The above spectral currents are subject to the following conservation law:
$\sum\limits_{\alpha=1}^{N_r}dI(E,t)_{\alpha\beta}/dE =0$. 
This property supports the point of view that these currents arise "inside" the scatterer
(they are generated by the nonstationary scatterer) 
without any external current source. 
Thus we can consider the pump as a source of currents rather then a
source of voltages \cite{RS03}.

For a scatterer with scattering matrix Eq.(\ref{Eq26}), 
we obtain the spectral currents
\begin{subequations}
\label{Eq35}
\begin{equation}
\label{Eq35A}
\frac{dI_{11}(E,t)}{dE}
= - \frac{e}{4\pi}\left(
\frac{\partial(\gamma - \theta)}{\partial t}\frac{\partial R}{\partial
E} -
\frac{\partial(\gamma - \theta)}{\partial E} \frac{\partial R}{\partial
t}
\right) ,
\end{equation}
\begin{equation}
\label{Eq35B}
\frac{dI_{22}(E,t)}{dE}
= - \frac{e}{4\pi}\left(
\frac{\partial(\gamma + \theta)}{\partial t}\frac{\partial R}{\partial
E} -
\frac{\partial(\gamma + \theta)}{\partial E} \frac{\partial R}{\partial
t}
\right).
\end{equation}
\end{subequations}
\noindent
Note that above currents depend on the phase $\gamma$  
related to the  charge of a scatterer.

Strictly speaking if we are dealing with time dependent currents 
(instead of only the time averaged currents) 
then we need to show that these currents satisfy the continuity 
equation for the charge currents:

\begin{equation}
\label{Eq36}
\sum\limits_{\alpha}I_{\alpha}(t) +  \frac{\partial Q(t)}{\partial t} =
0.
\end{equation}

\noindent
Here $I_{\alpha}(t)$ is the full time-dependent current flowing through
the scatterer
to the lead $\alpha$; $Q(t)$ is a charge of a scatterer.

To calculate $I_{\alpha}(t)$ we first calculate 
the Fourier transformed current 
$I_{\alpha,l} = \frac{\omega}{2\pi}\int\limits_{0}^{\cal T} dt
e^{il\omega t} I_{\alpha}(t)$
which reads \cite{Buttiker92}:
$$
I_{\alpha,l} = \frac{e}{h}\int\limits_{0}^{\infty}dE 
\left\{ \langle\hat b^{\dagger}_{\alpha}(E)\hat
b_{\alpha}(E+l\hbar\omega)\rangle 
- \langle\hat a^{\dagger}_{\alpha}(E)\hat
a_{\alpha}(E+l\hbar\omega)\rangle \right\}.
$$
The operators $\hat b_{\alpha}$ and $\hat a_{\alpha}$ are defined
in Eqs.(\ref{Eq8}) and (\ref{Eq9}), respectively. 
The calculations analogous to those leading to Eq.(\ref{Eq33}) give us 
$I_{\alpha,l}$.
Performing the inverse Fourier transformation Eq.(\ref{Eq14B}) we
finally get 
the time-dependent current $I_{\alpha}(t)$ flowing in the system as
follows:
\begin{widetext}
\begin{equation}
\label{Eq37} 
I_{\alpha}(t) =   
\int\limits_{0}^{\infty} dE 
\sum\limits_{\beta} 
\left\{  
\frac{e}{h}
\left[ f_{0}\big(E;\mu_{\beta}(t)\big) -
f_{0}\big(E;\mu_{\alpha}(t)\big) \right]  
\big|S_{0,\alpha\beta}(E,t)\big|^2 
- e \frac{\partial}{\partial t}
 \left[ f_{0}\big(E;\mu_{\beta}(t)\big)
\frac{dN_{\alpha\beta}(E,t)}{dE}\right]
+ f_{0}\big(E;\mu_{\beta}(t)\big) \frac{dI_{\alpha\beta}(E,t)}{dE}
\right\}. 
\end{equation}
\end{widetext}

\noindent
Here we have introduced the partial density of states \cite{BTP94}
for a "frozen" scatterer, 
$$
\frac{dN_{\alpha\beta}}{dE} = \frac{i}{4\pi}
\left( 
\frac{\partial S^{*}_{0,\alpha\beta}}{\partial E} S_{0,\alpha\beta}
- S^{*}_{0,\alpha\beta}\frac{\partial S_{0,\alpha\beta}}{\partial E} 
\right).
$$ 

These density of states define the charge $Q(t)$ of a "frozen" scatterer
as follows:
\begin{equation}
\label{Eq38}
Q(t) = e
\sum\limits_{\alpha} \sum\limits_{\beta} 
\int\limits_{0}^{\infty} dE f_{0}\big(E;\mu_{\beta}(t)\big)
\frac{dN_{\alpha\beta}(E,t)}{dE}
\end{equation}

The quantity $I_{\alpha}(t)$ Eq.(\ref{Eq37}) and $Q(t)$ Eq.(\ref{Eq38})
do satisfy the continuity equation (\ref{Eq36}).

The three terms in the curly brackets on the RHS of Eq.(\ref{Eq37})
can be interpreted as follows. 
The first term defines the currents flowing under the action of external
voltages
through a "frozen" scatterer.
The second one defines currents attributed  to an oscillating charge of
a "frozen" scatterer.
The third term can not be entirely viewed just as a nonadiabatic
correction of either 
the "frozen" conductance nor of the "frozen" density of
states. 
It is more naturally to consider it as the ac currents generated by the
oscillating scatterer.
The ability to generate these ac currents differentiates a nonstationary
dynamical scatterer from a merely "frozen" scatterer.

\section{Discussion}
\label{D} 

We have investigated the 
nonstationary adiabatic charge transport through a time-dependent
mesoscopic scatterer coupled to reservoirs subject to oscillating voltages. 
The external voltages applied to the  reservoirs 
induce ac currents flowing through the scatterer.
In addition the oscillating scatterer itself is a source of ac currents flowing
between the reservoirs. 
In general these two types of
currents interfere with themselves. This gives rise 
to renormalization of the rectification 
(i.e., proportional to the potential difference) 
contribution to the dc current and gives rise 
to a renormalization of the quantum pump current.

To analyze this interference effect we calculated the 
Floquet scattering matrix beyond 
the adiabatic approximation. 
We investigated the first order in $\omega$ corrections 
to the (adiabatic) scattering matrix
and found that the dc currents of both the zeroth 
and the first order in $\omega$ can be 
expressed in terms of a stationary scattering matrix 
with time-dependent parameters.
Within this approximation, within a non-interacting theory,  
the oscillating potentials 
$V_{\alpha}(t)$ of reservoirs can be accounted for
by allowing the energy $E$ of incident particles 
to follow adiabatically the reservoir
potential: $E\to {\cal E} =  E + eV_{\alpha}(t)$.

We emphasize the importance of the phases of all the 
cyclically evolving quantities
(the potentials of reservoirs and the parameters of a scatterer) 
for generating a dc current.
In particular even when all the reservoirs have 
the same oscillating potential 
$V_{\alpha}(t) = V(t)$ and the rectification effect is 
ineffective 
the dc currents at $V=0$ and at $V\neq 0$ can nevertheless 
differ significantly.

The analysis allows us to perform a current partition 
that clarifies the physical 
meaning of the (diagonal elements of the) 
quantity ${\cal P}\{\hat S_{0}^{\dagger};\hat S_{0}\}$
(the Floquet corrections to the "frozen scattering" matrix)
and show that they correspond to spectral current densities
generated by a dynamic scatterer. The instantaneous current 
contains a contribution from such self-generated ac currents 
in addition to the currents from the "frozen" charge and 
the ac currents generated by the external potentials. 

We emphasize that the results presented in this work, 
the effect of external ac-potentials on a quantum pump, 
are of importance whenever the pump is not part of an ideal 
zero-impedance external circuit. In particular, if the pump 
is in series with a resistance used to measure the voltage generated 
by the pump, or if the circuit is a multiterminal circuit with probes 
used to measure voltages, the results presented here will be needed.

\begin{acknowledgments}
This work is supported by the Swiss National Science Foundation.
\end{acknowledgments} 

\appendix*\section{}

\subsection{The matrix $\hat A$ }
\label{A1}

The matrix $\hat A$ defines the first order in frequency 
corrections to the adiabatic Floquet scattering matrix Eqs.(\ref{Eq17}),
(\ref{Eq20}). It is a matrix antisymmetric in lead indexes.

To obtain Eq.(\ref{Eq21}) 
we substitute the adiabatic expansion Eq.(\ref{Eq17A}) 
in the current conservation condition Eq.(\ref{Eq16A}).
Keeping terms of order $\omega^{0}$ and $\omega^{1}$ we get the
following:
$$
\sum\limits_{\alpha}\sum\limits_{n=-\infty}^{\infty}
S^{*}_{F,\alpha\beta}(E_{n},E)S_{F,\alpha\gamma}(E_{n},E_{m})   
$$
$$
\approx \sum\limits_{\alpha}\sum\limits_{n}
\left(S_{0,\alpha\beta,n}^{*}(E)
+ \hbar\omega\frac{n}{2}\frac{\partial 
S_{0,\alpha\beta,n}^{*}(E)}{\partial E}
+ \hbar\omega A_{\alpha\beta,n}^{*}(E)  \right) 
$$
$$
\times
\left(S_{0,\alpha\gamma,n-m}(E)
+ \hbar\omega\frac{n+m}{2}\frac{\partial 
S_{0,\alpha\gamma,n-m}(E)}{\partial E}
+ \hbar\omega A_{\alpha\gamma,n-m}(E) \right) 
$$
$$
\approx \sum\limits_{\alpha}\sum\limits_{n}
S^{*}_{0,\alpha\beta,n}(E)S_{0,\alpha\gamma,n-m}(E)  
$$
$$
+ \sum\limits_{\alpha}\sum\limits_{n}
S_{0,\alpha\beta,n}^{*}(E)
\left( \hbar\omega\frac{n+m}{2}\frac{\partial 
S_{0,\alpha\gamma,n-m}(E)}{\partial E}
+ \hbar\omega A_{\alpha\gamma,n-m}(E) \right) 
$$
$$
+\sum\limits_{\alpha}\sum\limits_{n}
 \left( \hbar\omega\frac{n}{2}\frac{\partial 
S_{0,\alpha\beta,n}^{*}(E)}{\partial E}
+ \hbar\omega A_{\alpha\beta,n}^{*}(E) \right) S_{0,\alpha\gamma,n-m}(E) 
$$
$$
= \delta_{m0}\delta_{\beta\gamma}.
$$
Applying the inverse Fourier transformation Eq.(\ref{Eq14B})
and introducing corresponding matrixes
we rewrite above equation as follows:

\begin{equation}
\label{EqA_1}
\begin{array}{c}
\Big(
\big|\hat S_{0}(E,t) \big|^2
+ \hbar\omega \hat S_{0}^{\dagger}(E,t)\hat A(E,t)
+ \hbar\omega \hat A^{\dagger}(E,t)\hat S_{0}(E,t)  \\
\ \\
+ \frac{i\hbar}{2} \hat S_{0}^{\dagger}\frac{\partial^2\hat 
S_{0}}{\partial E\partial t}
-  i\hbar\frac{\partial}{\partial t}\left[
\hat S^{\dagger}_{0} \frac{\partial \hat S_{0}}{\partial E} \right]
-  \frac{i\hbar}{2} \frac{\partial^2\hat  S_{0}^{\dagger}}{\partial
E\partial t} \hat S_{0}
\Big)_{\beta\gamma,-m}  = \delta_{m0}\delta_{\beta\gamma}.
\end{array}
\end{equation}
To simplify further this equation we use the unitarity condition for the
frozen 
scattering matrix: $\hat S_{0}(E,t)\hat S^{\dagger}_{0}(E,t) = \hat I$.
First, from this condition it follows that
\begin{equation}
\label{EqA_2}
\Big(\big|\hat S_{0}(E,t) \big|^2\Big)_{\beta\gamma,-m}  =
\delta_{m0}\delta_{\beta\gamma}.
\end{equation}
And second, we can write 
$\frac{\partial^2}{\partial E\partial t}\Big[\hat S_{0}^{\dagger}\hat
S_{0} \Big] = 0  $
and, correspondingly,
\begin{equation}
\label{EqA_3}
- \frac{\partial^2\hat S_{0}^{\dagger}}{\partial E\partial t}\hat S_{0}
- \hat S_{0}^{\dagger}\frac{\partial^2\hat S_{0}}{\partial E\partial t} 
=  \frac{\partial\hat S_{0}^{\dagger}}{\partial E}\frac{\partial\hat
S_{0}}{\partial t}
 + \frac{\partial\hat S_{0}^{\dagger}}{\partial t}\frac{\partial\hat
S_{0}}{\partial E}.
\end{equation}
Substituting Eqs.(\ref{EqA_2}) and (\ref{EqA_3}) in Eq.(\ref{EqA_1}) 
we arrive at the  Eq.(\ref{Eq21}).

\subsection{The commutator matrix ${\cal P}$}
\label{A2}
The matrix ${\cal P}\{\hat S_{0}^{\dagger};\hat S_{0} \} $ defined in
Eq.(\ref{Eq21B}) is self adjoint
\begin{equation}
\label{EqA_4}
{\cal P}\{\hat S_{0}^{\dagger};\hat S_{0} \} = 
{\cal P}^{\dagger}\{\hat S_{0}^{\dagger};\hat S_{0} \},
\end{equation}
\noindent
and traceless 
\begin{equation}
\label{EqA_5}
Tr\big[{\cal P}\{\hat S_{0}^{\dagger};\hat S_{0} \}\big] = 0. 
\end{equation}
To demonstrate the latter property we use the equality 
$d[\hat S]\hat S^{\dagger} = - \hat Sd[\hat S^{\dagger}]$
following from the unitarity of the scattering matrix
$\hat S\hat S^{\dagger} = \hat I$ 
and the invariance of trace to the cyclic
rearrangements of the matrices.
As a result from Eq.(\ref{Eq21B}) we get
$$
\begin{array}{c}
Tr[{\cal P}] = 
i\hbar Tr
\left[ \frac{\partial \hat S_{0}^{\dagger}}{\partial t}
\frac{\partial \hat S_{0}}{\partial E} -
\frac{\partial \hat S_{0}^{\dagger}}{\partial E}
\hat S\hat S^{\dagger}
\frac{\partial \hat S_{0}}{\partial t}
\right] \\
\ \\
=
i\hbar Tr
\left[ \frac{\partial \hat S_{0}^{\dagger}}{\partial t}
\frac{\partial \hat S_{0}}{\partial E} -
S^{\dagger}
\frac{\partial \hat S_{0}}{\partial E}
\frac{\partial \hat S_{0}^{\dagger}}{\partial t}
\hat S
\right] \\
\ \\
=
i\hbar Tr
\left[ \frac{\partial \hat S_{0}^{\dagger}}{\partial t}
\frac{\partial \hat S_{0}}{\partial E} -
\frac{\partial \hat S_{0}^{\dagger}}{\partial t}
\frac{\partial \hat S_{0}}{\partial E}
\right]  = 0 .
\end{array}
$$


\newpage

\begin{thebibliography}{11}

\bibitem{BTP94}
   M. B\"{u}ttiker, H. Thomas, and A. Pr\^{e}tre, 
   Z. Phys. B {\bf 94}, 133 (1994); 
   M. B\"{u}ttiker,
   J. Phys. Condensed Matter {\bf 5}, 9361 (1993).

\bibitem{PP94}
   J. B. Pieper and J.C. Price, 
   Phys. Rev. Lett. {\bf 72}, 3586 (1994).

\bibitem{TG63}
   P. K. Tien and J.P. Gordon, 
   Phys. Rev. {\bf 129}, 647 (1963).

\bibitem{TE73}
   R. Tsu and L. Esaki, 
   Appl. Phys. Lett. {\bf 22}, 562 (1973).

\bibitem{KJOENMS94}
   L. P. Kouwenhoven, S. Jauhar, J. Orenstein, P. L. McEuen, Y. Nagamune,
   J. Motohisa, and H. Sakaki, 
   Phys. Rev. Lett. {\bf 73}, 3443 (1994).

\bibitem{BHWKE95}
   R. H. Blick, R. J. Haug, D. W. van der Weide, K. von Klitzing, and  K. Eberl, 
   Appl. Phys. Lett. {\bf 67}, 3924 (1995).

\bibitem{SMCG99}
    M. Switkes, C. M. Marcus, K. Campman, and A. C. Gossard,
    Science {\bf 283}, 1905 (1999).

\bibitem{MSKNJU02}
    R. G. Mani, J. H. Smet, K. von Klitzing, V. Narayanamurti, W. B. Johnson, and V. Umansky,
    Nature, {\bf 420}, 646 (2002).

\bibitem{ZDPW03}
    M. A.  Zudov, R. R. Du, L. N. Pfeiffer, and K. W. West, 
    Phys. Rev. Lett. {\bf 90}, 046807 (2003).

\bibitem{RRGEJ03}
   L.-H. Reydellet, P. Roche, D.C. Glattli, B. Etienne, and Y. Jin, 
   Phys. Rev. Lett. {\bf 90}, 176803 (2003).

\bibitem{WPMU03}
    S. K. Watson, R. M. Potok, C. M. Marcus, and V. Umansky,
    cond-mat/0302492, (unpublished).

\bibitem{DMH03}
    L. DiCarlo, C.M. Marcus, and J.S. Harris J.,
    cond-mat/0304397, (unpublished).

\bibitem{Brouwer98}
    P. W. Brouwer, 
    Phys. Rev. B {\bf 58}, R10135 (1998).

\bibitem{AA98}
   I.L. Aleiner and A.V. Andreev, 
   Phys. Rev. Lett. {\bf 81}, 1286 (1998).

\bibitem{ZSA99}
    F. Zhou, B. Spivak, and B. Altshuler, 
    Phys. Rev. Lett. {\bf 82}, 608  (1999).

\bibitem{WS99}
     M. Wagner and F.Sols, 
     Phys. Rev. Lett. {\bf 83}, 4377 (1999).

\bibitem{AK00}
    A. Andreev and A. Kamenev,
    Phys. Rev. Lett. {\bf 85}, 1294 (2000).

\bibitem{SAA00}
    T. A. Shutenko, I. L. Aleiner, and B. L. Altshuler, 
    Phys. Rev. B {\bf 61}, 10366 (2000).

\bibitem{WWG00}
    Y. Wei, J. Wang, and H. Guo, 
    Phys. Rev. B {\bf 62}, 9947 (2000).

\bibitem{AEGS00}
    J. E. Avron, A. Elgart, G. M. Graf, and L. Sadun, 
    Phys. Rev. B {\bf 62}, 10618 (2000).

\bibitem{VAA01}
    M. G. Vavilov, V. Ambegaokar, and I. L. Aleiner,
    Phys. Rev. B {\bf 63}, 195313 (2001).

\bibitem{AEGS01}
    J. E. Avron, A. Elgart, G. M. Graf, and L. Sadun,
    Phys. Rev. Lett. {\bf 87}, 236601 (2001);
    J. Math. Phys. 43: 3415 (2002). 

\bibitem{TC01}
    C.-S. Tang and C.S. Chu, 
    Solid State  Commun. {\bf 120}, 353 (2001).

\bibitem{ZW02}
    S.-L. Zhu and Z.D. Wang, 
    Phys. Rev. B {\bf 65}, 155313 (2002).

\bibitem{EWAL02}
    O. Entin-Wohlman, A. Aharony, and Y. Levinson, 
    Phys. Rev. B {\bf 65}, 195411 (2002).

\bibitem{MB02}
    M. Moskalets and M. B\"{u}ttiker, 
    Phys. Rev. B {\bf  66}, 035306  (2002).

\bibitem{WW02}  
    B. Wang and J. Wang,  
    Phys. Rev. B {\bf 66}, 125310 (2002).

\bibitem{Kim02}  
    S. W. Kim,   
    Phys. Rev. B {\bf 66}, 235304 (2002);

\bibitem{PVB02}    
    M. L. Polianski, M. G. Vavilov, and P. W. Brouwer,  
    Phys. Rev. B {\bf 65}, 245314 (2002).  

\bibitem{MBstrong02}
    M. Moskalets and M. B\"{u}ttiker, 
    Phys.Rev. B {\bf 66}, 205320 (2002).

\bibitem{PB02}    
    M. L. Polianski and P. W. Brouwer,  
    cond-mat/0208408 (unpublished). 

\bibitem{GTF02}
    M. Governale, F. Taddei, and R. Fazio, 
    cond-mat/0211211 (unpublished). 

\bibitem{WWG02}
    B. Wang, J. Wang, and H. Guo, 
    cond-mat/0211536 (unpublished). 

\bibitem{SC02}
    P. Sharma and C. Chamon, 
    cond-mat/0212201 (unpublished). 

\bibitem{Aono03}
    T. Aono, 
    Phys. Rev. B. {\bf 67}, 155303 (2003).

\bibitem{GTM03}
    V. Gudmundsson, C.-S. Tang, and  A. Manolescu, 
    Phys. Rev. B {\bf 67}, 161301 (2003).

\bibitem{MBhidden03} 
    M. Moskalets and M. B\"{u}ttiker, 
    Phys.Rev. B {\bf 68}, 075303 (2003).

\bibitem{ZCMcK03}
    H.-Q. Zhou, S.Y. Cho, and R.H. McKenzie, 
    cond-mat/0304205 (unpublished);
    H.-Q. Zhou, U. Lundin, S. Y. Cho and R. H. McKenzie, 
    cond-mat/0309096 (unpublished).
   
\bibitem{Cohen03}
    D. Cohen, 
    cond-mat/0304678 (unpublished).

\bibitem{AEGS03}
    J.E. Avron, A. Elgart, G.M. Graf, and L. Sadum, 
    math-ph/0305049 (unpublished).

\bibitem{CTCC03}
    S.W. Chung, C.-S. Tang, C.S. Chu, and C.Y. Chang,  
    cond-mat/0306194 (unpublished).

\bibitem{Blaauboer03}
     M. Blaauboer,  
     cond-mat/0307166, (unpublished); 
     cond-mat/0309496 (unpublished).

\bibitem{BDR03}
    A. Banerjee, S. Das and S. Rao,   
    cond-mat/0307324 (unpublished).

\bibitem{RS03}
    M. Rey and F. Sols, 
    cond-mat/0308257 (unpublished).

\bibitem{KAEW03}
    V. Kashcheyevs, A. Aharony, and O. Entin-Wohlman,  
    cond-mat/0308382  (unpublished).

\bibitem{EWAK03}
    O. Entin-Wohlman, A. Aharony, and V. Kashcheyevs,  
    cond-mat/0308408  (unpublished).

\bibitem{Brouwer01}
    P. W. Brouwer, 
    Phys. Rev. B {\bf 63}, 121303 (2001).

\bibitem{PB01}
    M. L. Polianski and P. W. Brouwer, 
    Phys. Rev. B {\bf 64}, 075304 (2001).

\bibitem{MB01}
    M. Moskalets and M. B\"{u}ttiker, 
    Phys. Rev. B. {\bf 64}, 201305 (2001).
    
\bibitem{MMLM}
   M. Martinez-Mares, C. H. Lewenkopf, and E. R. Mucciolo, 
   cond-mat/0309197 (unpublished). 

\bibitem{Buttiker90}
   M. B\"{u}ttiker, 
   Phys. Rev. Lett. {\bf 65}, 2901 (1990).

\bibitem{Buttiker92}
   M. B\"{u}ttiker, 
   Phys. Rev. B, {\bf 46}, 12485 (1992).

\bibitem{PB98}   
   M. H. Pedersen and M. B\"uttiker, 
   Phys. Rev. B {\bf58}, 12993 (1998).

\bibitem{Wigner55} 
   E.P.  Wigner, 
   Phys. Rev. {\bf 98}, 145 (1955).

\bibitem{Smith60} 
   F.T. Smith, 
   Phys. Rev. {\bf 118}, 349 (1960).

\bibitem{Bateman} 
    H. Bateman, 
    Higher transcendental functions. Ed. by A.Erd\'elyi, New York Toronto London, 
    Mc Graw-Hill Book Company, inc, (1953). 

\bibitem{Yafaev92}
    D.R. Yafaev, 
    Mathematical scattering theory, AMS (1992).

\bibitem{Friedel52}
   J. Friedel, 
   Phil. Mag. {\bf 43}, 153 (1952).

\end{thebibliography}
\end{document}